\documentclass[12pt,reqno]{amsart}
\usepackage{epsfig}
\usepackage{latexsym}

\newcommand{\szego}{Szeg\"o }

\newcommand{\nbar}{\raisebox{2pt}{$\bar{\ }$}}

\newcommand{\inv}{^{-1}}
\newcommand{\kahler}{K\"ahler }
\newcommand{\sqrtn}{\sqrt{N}}
\newcommand{\wt}{\widetilde}
\newcommand{\wh}{\widehat}
\newcommand{\PP}{{\mathbb P}}

\newcommand{\C}{{\mathbb C}}
\newcommand{\Q}{{\mathbb Q}}

\newcommand{\CP}{\C\PP}
\renewcommand{\d}{\partial}

\renewcommand{\H}{{\mathbf H}}
\newcommand{\half}{{\frac{1}{2}}}

\newcommand{\SU}{{\operatorname{SU}}}

\renewcommand{\phi}{\varphi}

\newcommand{\go}{\mathfrak}

\newcommand{\ccal}{\mathcal{C}}

\newcommand{\ical}{\mathcal{I}}
\newcommand{\lcal}{\mathcal{L}}
\newcommand{\mcal}{\mathcal{M}}

\newcommand{\scal}{\mathcal{S}}

\newcommand{\al}{\alpha}
\newcommand{\be}{\beta}
\newcommand{\ga}{\gamma}

\newcommand{\La}{\Lambda}

\newcommand{\de}{\delta}
\newcommand{\De}{\Delta}
\newcommand{\om}{\omega}
\newcommand{\Om}{\Omega}

\newtheorem{theo}{{\sc Theorem}}[section]
\newtheorem{cor}[theo]{{\sc Corollary}}
\newtheorem{lem}[theo]{{\sc Lemma}}
\newtheorem{prop}[theo]{{\sc Proposition}}

\newenvironment{rem}{\medskip\noindent{\it Remark:\/} }{\medskip}

\title[Correlations between zeros and supersymmetry]{Correlations
between zeros and supersymmetry}

\author{Pavel Bleher}
\address{Department of Mathematical Sciences, IUPUI, Indianapolis, IN
46202,
USA}
\email{bleher@math.iupui.edu}
\author{Bernard Shiffman}
\address{Department of Mathematics, Johns Hopkins University, Baltimore,
MD
21218, USA}
\email{shiffman@math.jhu.edu}
\author{Steve Zelditch}
\address{Department of Mathematics, Johns Hopkins University, Baltimore,
MD
21218, USA}
\email{zelditch@math.jhu.edu}

\thanks{Research partially supported by NSF grants
\#DMS-9970625 (first author), \#DMS-9800479 (second author),
\#DMS-0071358(third author).}

\date{November 19, 2000}

\begin{document}

\begin{abstract} 
In our previous work \cite{BSZ2}, we proved that the correlation functions for
simultaneous zeros of random generalized polynomials have universal scaling
limits and we gave explicit formulas for pair correlations in codimensions 1 and
2. The purpose of this paper is to compute these universal limits in all
dimensions and codimensions.  First, we use a supersymmetry method to express
the $n$-point correlations as Berezin integrals.  Then we use the
Wick method to give a closed formula for the limit pair correlation function for
the point case in all dimensions.
\end{abstract}

\maketitle

\section{Introduction}  This paper is a continuation of our   articles \cite{BSZ1,
BSZ2, BSZ3} on the correlations between zeros of random
holomorphic polynomials in $m$ complex variables  and their generalization to
holomorphic sections of positive line bundles $L\to M$ over general \kahler
manifolds of dimension $m$ and their symplectic counterparts.  These correlations
are defined by the probability density 
$K^N_{nk}(z^1,\dots,z^n)$ of finding joint zeros of $k$ independent sections at
the points $z^1, \dots, z^n\in M$  (see \S 2). 
 To obtain universal quantities, we  re-scale the correlation functions in
normal coordinates by a factor of $\sqrtn$. Our main result from \cite{BSZ2,
BSZ3} is that the (normalized) correlation functions have a universal scaling
limit
\begin{equation}\label{slcd}\wt K^\infty_{nkm}(z^1,\dots,z^n)= \lim
_{N\to\infty}K^N_{1k}(z_0)^{-n}
K_{nk}^N(z_0+\frac{z^1}{\sqrtn},\dots,z_0+\frac{z^n}{\sqrtn})\,,\end{equation} 
which is independent of the manifold $M$, the line bundle $L$ and the point
$z_0$;
$\wt K^\infty_{nkm}$ depends only on the dimension
$m$ of the manifold and the codimension $k$ of the zero set.  The problem then arises of calculating these universal
functions explicitly and analyzing their small distance and large distance
behavior.  In \cite{BSZ1}, \cite{BSZ2}, we gave explicit formulas for the pair
correlation functions $\wt K^\infty_{2km}(z^1,z^2)$ in codimensions $k = 1,2$,
respectively.   The purpose of this paper is to complete these results by  giving explicit   formulas for $\wt
K^\infty_{nkm}$ in all dimensions and codimensions.

 Our 
first formula expresses the correlation
as  a supersymmetric (Berezin) integral involving the matrices
$\La(z),\ A^\infty(z)$ used in our prior formulas, as well as a  matrix
$\Om$ of fermionic variables described below.
\begin{theo} \label{susy} The limit $n$-point correlation functions are given by
$$\wt K^\infty_{nkm}(z^1,\dots,z^n)=
\frac{[(m-k)!]^n}{(m!)^{n}[\det A^\infty(z)]^k}\int
\frac{1}{\det[I+\La(z)\Om]}d\eta \,.$$
\end{theo}
\noindent Here, $\Om$ is the $nkm\times nkm$ matrix
\begin{equation}\label{Omega}\Om=\left(
\Om_{p'j'q'}^{pjq}\right)= \left( \de^p_{p'}\de^q_{q'}
\eta^{p'}_{j'}\bar\eta^p_j
\right)\qquad (1\le p,p'\le n,\ 1\le j,j'\le k,\ 1\le q,q'\le m)\,,\end{equation}
where the $\eta_j,\bar\eta_j$ are anti-commuting (fermionic) variables, and
$d\eta=\prod_{j,p}d\bar\eta_j^p d\eta_j^p$.   The integral in Theorem \ref{susy}
is a Berezin integral, which is evaluated by simply taking the coefficient of
the top degree form of the integrand
${\det[I+\La(z)\Om]}\inv$ 
(see \S \ref{berezin}).  Hence the formula in Theorem \ref{susy} is a purely
algebraic expression in the coefficients of
$\La(z)$ and $A^\infty(z)$, which are given
in terms of the \szego kernel of the Heisenberg group and its derivatives (see
\S
\ref{review}).  We remark that supersymmetric methods have also been applied to limit
correlations in random matrix theory by Zirnbauer \cite{Zi}.

In  the case $n = 2$,   $\wt K_{2km}^\infty(z^1,z^2)$, depends only on the
distance between the points $z^1,z^2$, since it is universal and hence invariant
under rigid motions.  Hence it may be written as:
\begin{equation} \wt K_{2km}^\infty(z^1,z^2) = \kappa_{km}(|z^1-z^2|)\,.
\end{equation}
We refer to \cite{BSZ2} for details. In  \cite{BSZ1} we gave an explicit formula for 
$\kappa_{1m}$  (using the ``Poincar\'e-Lelong
formula"), and in \cite{BSZ2} we evaluated $\kappa_{2m}$. (The  pair correlation function $\kappa_{11}(r)$  was first 
determined by Hannay \cite{Ha} in 
the case of zeros of $SU(2)$ polynomials in one complex variable.)
In \S
\ref{pair}, we use Theorem \ref{susy} to give the following new  Berezin
integral formula for
$\kappa_{km}$:

\begin{cor} The pair correlation functions are given by
$$\kappa_{km}(r)=\frac{(m-k)!^2}{m!^2(1-e^{-r^2})^k}
\int\frac{1}{\Phi\Psi^{m-1}}d\eta\,,$$
where
\begin{eqnarray*}\Phi &=&
\det\left[I+P(\Om_1+\Om_2)+T\Om_1\Om_2\right]\,,\\
&&P\ =\ 1-\frac{r^2e^{-r^2}}{1-e^{-r^2}}\,,\quad T\ = \ 
1-e^{-r^2}-\frac{r^4e^{-r^2}}{1-e^{-r^2}}\,,\\[8pt]
\Psi &=& 
\det\left[I+\Om_1+\Om_2+(1-e^{-r^2})\Om_1\Om_2\right]
\,.\end{eqnarray*}
\label{susy2}\end{cor}
\noindent Here, $\Om_1,\ \Om_2$ are the $k\times k$ matrices
$$\Om_p=\left( \eta^{p}_{j'}\bar\eta^p_j
\right)_{1\le j,j'\le k}\,,\qquad p=1,2\,.$$ We then expand  the formula
as a (finite) series (\ref{km}), which we use to compute explicit formulas
for
$\kappa_{km}$. 

The most
vivid
case is when $k =m$, where the simultaneous zeros of
$k$-tuples of sections almost surely form a set of discrete points.  Our
second result is an explicit formula  for the point pair correlation functions
$\kappa_{mm}$ in all dimensions:

\begin{theo} \label{2mm} The point pair correlation functions are given by
\begin{equation}\begin{array}{r}\kappa_{mm}(r) = \frac{
m (1-{v}^{m+1} ) (1-v )+r^2(2m+2)
 ({v}^{m+1}-v )+r^4\left[{v}^{m+1}+{v}^{m}+ ( \{m+1
\}v+1 )(v^m-v)/(v-1)\right]}
{m (1-v
)^{m+2}}\,,\\[8pt]  v=e^{-r^2}\,,
\end{array}\label{pointpaircor}\end{equation}
for $m\ge 1$.
For small values of $r$, we have
\begin{equation} \label{leading} \kappa_{mm}(r)= \frac{m+1}{4}
r^{4-2m} + O(r^{8-2m})\,,\qquad\mbox{as }\ r\to 0\,.\end{equation}
\end{theo}

We prove Theorem \ref{2mm} in \S \ref{point} without making use of supersymmetry. 
Our proof uses instead the Wick formula expansion of the Gaussian integral
representation of the correlation. 

It is interesting to observe  the dimensional dependence of the short distance
behavior of $\kappa_{mm}(r)$. When
$m = 1, \kappa_{mm}(r) \to 0$ as $r \to 0$ and one has ``zero repulsion.''
When $m = 2$, $\kappa_{mm}(r) \to  3/4$ as $r \to 0$ and one has a kind of
neutrality. With $m \geq 3$, $\kappa_{mm}(r)
\nearrow \infty$ as $r \to 0$ and there is some kind of attraction between zeros. 
More precisely, in dimensions greater than 2, one is more likely to find a zero at a
small distance
$r$ from another zero than at a small distance $r$ from a given point; i.e.,  
zeros tend to clump together in high dimensions.  Indeed, in all dimensions,
the probability of finding another zero in a ball of small scaled radius $r$
about another zero is $\sim r^4$. We give below a graph of $\kappa_{33}$; graphs of
$\kappa_{11}$ and $\kappa_{22}$ can be found in \cite{BSZ2}.

\ \vspace{-.8in}
\begin{figure}[ht]\label{kappa33}\centering
\epsfig{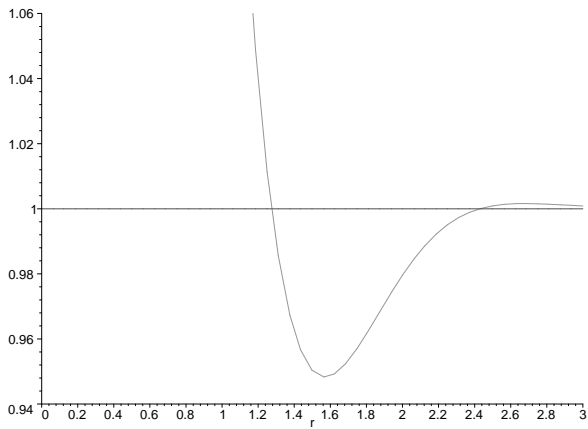} 
\caption{The limit pair correlation function
$\kappa_{33}$}
\end{figure}

\begin{rem} Theorem \ref{2mm} says that the {\it expected number\/} of zeros in
the punctured ball of scaled radius $r$ about a given zero is $\sim \int_0^r
\kappa_{mm}(t) t^{2m-1}dt \sim r^4$.  But, one can show that for
balls of small scaled radii $r$, the expected number of zeros approximates the
probability of finding a zero.
\end{rem}

\section {Background} \label{review} We begin by recalling the
 scaling limit zero correlation formula of
\cite{BSZ2}.  Consider a random polynomial $s$ of degree $N$ in $m$
variables.  More generally, $s$ can be a random section of the
$N^{\rm th}$ power $L^N$ of a positive line bundle $L$ on an $m$-dimensional
compact complex manifold $M$ (or a symplectic $2m$-manifold; see
{\cite{SZ,BSZ3}). We give $M$ the \kahler metric induced by the curvature form
$\om$ of the line bundle
$L$.  The probability measure on the
space of sections is the complex Gaussian measure induced by the Hermitian
inner product
$$\label{inner}\langle s_1, \bar s_2 \rangle = \int_M h^N(s_1,\bar
s_2)dV_M\;,$$ where $h^N$ is the metric on $L^N$ and 
$dV_M$ is the volume measure induced by $\om$.  (For further discussion of the
topics of this section, see
\cite{BSZ2}.) In particular, if $L$ is the hyperplane section bundle over
$\CP^m$, then random sections of $L^N$ are
polynomials of degree
$N$ in
$m$ variables of
the form
$$
P(z_1,\ldots,z_m)=\sum_{|J|\le N}\frac{C_J}{\sqrt{(N-|J|)!j_1!\cdots j_m!}}
z_1^{j_1}\cdots z_m^{j_m}\qquad (\,J=(j_1,\dots,j_m)\,)\;,$$ where the $C_J$ are i.i.d.\
Gaussian random variables with mean 0; they are called 
``$\SU(m+1)$-polynomials."

We consider  $k$-tuples $s=(s_1,\dots,s_k)$ of  i.i.d.\ random polynomials (or
sections) $s_j$ ($1\le k\le m$). 
The zero correlation density $K^N_{nk}(z^1,\dots,z^n)$ is defined as
the expected joint volume density of zeros of sections of $L^N$ at the points
$z^1,\dots,z^n$. In the case $k=m$, where the zero sets are discrete points, 
$K^N_{nk}(z^1,\dots,z^n)$ can be interpreted as the probability density of
finding simultaneous zeros at these points.   For instance, the zero density
function
$K^N_{1k}(z)\approx c_k N^k$ as $N\to\infty$, where
$c_k$ is independent of the point $z$ (see \cite{SZ1}).

In 
\cite{BSZ2,BSZ3}, we gave  generalized forms of the Kac-Rice formula
\cite{Ka,Ri}, which we used to express $K^N_{nk}(z^1,\dots,z^n)$  in terms of
the joint probability distribution (JPD) of the random variables
$s(z^1),\dots,s(z^n), \nabla s(z^1),\dots,\nabla s(z^n)$.  We then showed that
the scaling limit correlation function $\wt K^\infty_{nkm}$ given
 by (\ref{slcd}) can be expressed in terms of the scaling limit of the JPD.  

The central result of \cite{BSZ2} is that the limit JPD is universal and
can be expressed in terms of the \szego kernel $\Pi_1^\H$ for the Heisenberg
group:
\begin{equation}\label{heisenberg}\Pi_1^\H(z,\theta;w,\phi)
= \frac{1}{\pi^m}e^{i(\theta-\phi +\Im z\cdot \bar w)-\half|z-w|^2}
= \frac{1}{\pi^m} e^{i(\theta-\phi)+z\cdot
\bar
w-\half(|z|^2+|w|^2)}\,.
\end{equation}
To be precise, the limit JPD is a complex Gaussian measure with covariance
matrix
$\De^\infty$ given by:
\begin{equation}\label{delta}
\De^\infty(z)= \frac{m!}{\pi^m }\left(
\begin{array}{cc}
A^\infty(z) & B^\infty(z) \\
B^{\infty}(z)^* & C^\infty(z)
\end{array}\right)\,,\end{equation} where
\begin{eqnarray}
 \pi^{-m } A^\infty(z)^p_{p'} &=&
\Pi_1^\H(z^p,0;z^{p'},0)\,,\nonumber \\ \pi^{-m }
B^\infty(z)^{p}_{p'q'} &=& \frac 
{\nabla}{\d \bar z^{p'}_{q'}}\Pi_1^\H(z^p,0;z^{p'},0) \ = \
(z^p_{q'}-z^{p'}_{q'})
\Pi_1^\H(z^p,0;z^{p'},0)
\ ,\label{ABC}\\ \pi^{-m }C^\infty(z)^{pq}_{p'q'}
&=& \frac 
{\nabla^2}{\d  z^{p}_{q}\d \bar z^{p'}_{q'}}\Pi_1^\H(z^p,0;z^{p'},0)
\ = \
\left(\delta_{qq'}+(\bar z^{p'}_q -\bar z^p_q)
(z^p_{q'}-z^{p'}_{q'})\right)\Pi_1^\H(z^p,0;z^{p'},0) \  .
\nonumber \end{eqnarray}
(Here
$A^\infty,\ B^\infty,\ C^\infty$ are $n\times n,\ n\times mn,\
mn\times mn$ matrices, respectively.)  In the sequel, we shall use the matrix
\begin{equation}\label{La}
\La^\infty(z): = C^\infty(z) -B^\infty(z)^* A^\infty(z)\inv
B^\infty(z)\,.
\end{equation}  We note that $A^\infty(z)$ and $\La^\infty(z)$ are positive
definite whenever $z^1,\dots,z^n$ are distinct points. 

In
\cite{BSZ2}, we gave the following key formula for the limit correlation
functions:

\begin{equation}\label{Knkm}\wt K^\infty_{nkm}(z^1,\dots,z^n)=
\frac{[(m-k)!]^n}{(m!)^{n}[\det A^\infty(z)]^k}\int_{\C^{kmn}} \prod_{p=1}^n
\det_{1\le j,j'\le k}\left(
\sum_{q=1}^m \xi_{jq}^p \bar \xi_{j'q}^p
\right)d\ga_{\La(z)}(\xi)\,,\end{equation}
where $\ga_{\La(z)}$ is the Gaussian measure
with ($nkm\times nkm$) covariance matrix
\begin{equation}\label{Lambda2}\La(z):=\left(
\La(z)_{p'j'q'}^{pjq}\right)=\left(
\de^j_{j'}
\La^\infty(z)^{pq}_{p'q'}
\right)\,.\end{equation}  
(I.e., $\langle  \xi_{jq}^p \bar
\xi_{j'q'}^{p'}\rangle_{\ga_{\La(z)}}=\La(z)_{p'j'q'}^{pjq}$.)
For the pair correlation case ($n=2$),  equation (\ref{Knkm}) becomes:
\begin{equation}\label{slpc} \kappa_{km}(r) =
\frac{1}{\big[\frac{m!}{(m-k)!}\big]^2
\det A(r)^k}\int_{\C^{2km}} 
\det_{1\le j,j'\le k}\left(
\sum_{q=1}^m \xi_{jq}^1 \bar \xi_{j'q}^1
\right)
\det_{1\le j,j'\le k}\left(
\sum_{q=1}^m \xi_{jq}^2 \bar \xi_{j'q}^2
\right)d\ga_{\La(r)}(\xi)\,,\end{equation} 
where $$A(r)=A^\infty(z^1,z^2)\,,\ 
\La(r)=\La(z^1,z^2)\,,\quad |z^1-z^2|=r\,.$$

The computations in this paper are all
based on formula   (\ref{Knkm}).

\section{Supersymmetric approach to $n$-point correlations} \label{berezin}

We now prove Theorem \ref{susy} using our formula (\ref{Knkm}) for the limit
$n$-point correlation function, which we restate as follows:
\begin{equation}\label{Knkm+}\wt K^\infty_{nkm}(z^1,\dots,z^n)=
\frac{[(m-k)!]^n}{(m!)^{n}[\det A^\infty(z)]^k}G_{nkm}\,,\end{equation}
where
\begin{equation}\label{Gnkm} G_{nkm}(z)=
\int_{\C^{kmn}} \prod_{p=1}^n
\det_{1\le j,j'\le k}\left(
\sum_{q=1}^m \xi_{jq}^p \bar \xi_{j'q}^p
\right)d\ga_{\La(z)}(\xi)\,.\end{equation}
Our approach is to represent the determinant in (\ref{Gnkm}) as a Berezin
integral and then to exchange the order of integration.

We introduce anti-commuting (or ``fermionic") variables
$\eta_j^p,\bar\eta_j^p$ ($1\le j\le k,\; 1\le p\le n$), which
can be regarded as generators of the Grassmann algebra
$\bigwedge^\bullet
\C^{2l}=\bigoplus_{t=0}^{2l}\bigwedge^t
\C^{2l}$, $l=nk$.  The {\it Berezin integral\/} on
$\bigwedge^\bullet
\C^{2l}$ is the linear functional $\ical:\bigwedge^\bullet
\C^{2l}\to \C$ given by $$\ical|_{\bigwedge^t
\C^{2l}}=0\quad \mbox{for \ } t<2l\,,\quad \textstyle
\ical\left(\prod_{j,p}\eta_j^p \bar\eta_j^p\right)=1\,.$$
Elements $f\in \bigwedge^\bullet
\C^{2l}$ are considered as functions of anti-commuting
variables, and we write
$$\ical(f)=\int fd\eta=\int f\, \textstyle\prod_{j,p}d\bar\eta_j^p
d\eta_j^p\,.$$ (See for example
\cite[Chapter~2]{Ef}, \cite[\S 2.1]{ID}.)
If $H=\left(H^{pj}_{p'j'}\right)$ is an $l\times l$ complex matrix, we
have the supersymmetric formula for the determinant:
\begin{equation}\label{susy-det} \det H= \int e^{-\langle
H\eta,\bar\eta\rangle} d\eta\,,\qquad  \langle
H\eta,\bar\eta\rangle = \sum_{j,p,j',p'}\eta_j^p H^{pj}_{p'j'} \bar\eta_{j'}^{p'}
\,.\end{equation}

We now use (\ref{susy-det}) to compute $G_{nkm}$: let
$$\xi^p=\left(\begin{array}{ccc}\xi^p_{11} &
\cdots  & \xi^p_{1m}\\ \vdots &  & \vdots\\ \xi^p_{k1} &
\cdots  & \xi^p_{km}
\end{array}\right)$$ (where $\{\xi^p_{jq}\}$ are
ordinary
``bosonic" variables). We also write
$\xi=\xi^1\oplus\cdots\oplus\xi^n:\C^{mn}\to\C^{kn}$. Then
\begin{equation}  \prod_{p=1}^n
\det_{1\le j,j'\le k}\left(
\sum_{q=1}^m \xi_{jq}^p \bar \xi_{j'q}^p
\right) = \det (\xi\xi^*) = \det \left(\begin{array}{ccc}\xi^1\xi^{1*} &
\cdots  & 0\\ \vdots & \ddots & \vdots\\ 0 & \cdots & \xi^n\xi^{n*}
\end{array}\right)\,.\end{equation}
Applying (\ref{susy-det}) with $H=\xi\xi^*$, we have
\begin{eqnarray} G_{nkm} &=& \frac{1}{\pi^{nkm}\det\La}\int_{\C^{nkm}}
\det(\xi\xi^*)e^{-\langle \La\inv\xi,\bar\xi\rangle}\ d\xi\nonumber \\
&=& \frac{1}{\pi^{nkm}\det\La}\int_{\C^{nkm}}\int
e^{-\langle \La\inv\xi,\bar\xi\rangle-\langle \xi\xi^*\eta,\bar\eta\rangle}\
d\eta d\xi
\,,\label{expandG}\\
\langle \xi\xi^*\eta,\bar\eta\rangle &=&
\sum_{p,q,j,j'} \xi^p_{jq}\bar\xi^p_{j'q}\eta^p_{j'} \bar\eta^p_j=
\langle\Om\xi,\bar\xi\rangle\,,\label{expandOmega}
\end{eqnarray}
where $\Om$ is given by (\ref{Omega}).  Note that the entries of $\Om$ commute,
since they are of degree 2. Furthermore, adopting the supersymmetric
definition of the conjugate \cite{Ef}, $$(\eta^p_j)\nbar=\bar\eta^p_j,\quad
(\bar\eta^p_j)\nbar=-\eta^p_j\,,$$
we see that
the matrix $\Om$ is superhermitian; i.e., 
$\Om^*=\Om$, where $\Om^*={}^t\Om\nbar$.

Thus by (\ref{expandG})--(\ref{expandOmega}), we have
\begin{equation} G_{nkm}=\frac{1}{\pi^{nkm}\det\La}
\int_{\C^{nkm}}\int e^{-\left\langle (\La\inv+\Om)\xi,\bar\xi\right\rangle}\
d\eta d\xi \label{doubleint}
\,.\end{equation}
We recall that
\begin{equation}\frac{1}{\pi^{nkm}}\int_{\C^{nkm}}e^{-\langle
P\xi,\bar\xi\rangle}d\xi =
\det P\inv\,,\label{gaussint}\end{equation} for a positive definite, Hermitian
($nkm\times nkm$) matrix $P$. Furthermore, 
(\ref{gaussint}) holds when
$P$ is the superhermitian matrix $\La\inv+\Om$; we give a short proof of this
fact below.  Reversing the order of integration in (\ref{doubleint}) and
applying (\ref{gaussint}) with $P=\La\inv+\Om$, we have 
\begin{eqnarray} G_{nkm}&=&\frac{1}{\det\La}\int \frac{1}{\det(\La\inv
+\Om)}d\eta\nonumber\\ &=& \int \frac{1}{\det(I+\La\Om)}d\eta \,.
\label{susy0}\end{eqnarray}

We now verify by
formal substitution that (\ref{gaussint}) holds when  $P=\La\inv+\Om$:  
Suppose that $\Theta_1,\cdots\Theta_t$ are even fermionic functions; i.e.,
$\Theta_j\in 
\bigwedge^{\rm even}\C^{2l}$.  Let $S_\Theta$ be the homomorphism from
the algebra
$\C\{z_1,\dots,z_t,\bar z_1,\dots,\bar z_t\}$  of convergent power series onto
$\bigwedge^{\rm even}\C^{2l}$  given by substitution:
$$S_\Theta[f(z_1,\dots,z_t,\bar z_1,\dots,\bar
z_t)]=f(\Theta_1,\dots,\Theta_t,\bar\Theta_1,\dots,\bar\Theta_t)\,.$$ Let 
\begin{equation} f(Z,\bar Z)
= \int e^{-\langle(\La\inv+Z+Z^*)\xi,\bar\xi\rangle}d\xi = 
\int e^{-\langle(Z+ Z^*)\xi,\bar\xi\rangle}
e^{-\langle \La\inv\xi,\bar\xi\rangle}d\xi =
\frac{1}{\det(\La\inv+Z+Z^*)}\,,\label{formal}
\end{equation}
for $Z=(z_{\al\be})\in {\go
gl}(nkm,\C)$, where the last equality is by (\ref{gaussint}).  We easily see
that $f$ is a convergent power series in $\{z_{\al\be},\bar z_{\al\be}\}$
and that the integrand  $e^{-\langle(Z+ Z^*)\xi,\bar\xi\rangle}$ can be written as
an absolutely convergent power
series in $\{z_{\al\be},\bar z_{\al\be}\}$ with values in
$\lcal^1( e^{-\langle \La\inv\xi,\bar\xi\rangle}d\xi)$. We now let
$\Theta_{\al\be}=\half\Om_{\al\be}=\half\bar\Om_{\be\al}$; the conclusion follows
by applying $S_\Theta$ to (\ref{formal}). \qed

\subsection{Pair correlation}  \label{pair}  In this section, we prove
Corollary
\ref{susy2}.  To
illustrate the computation, we consider first the  case $k=m=1$ of zero
correlations in dimension one:  We have 
$$\La=\left(\begin{array}{cc}\La^1_1 & \La^1_2\\\La^2_1 &
\La^2_2\end{array}\right)\,,\quad \Omega = 
\left(\begin{array}{cc}\eta^1 \bar\eta^1 & 0\\0 &
\eta^2\bar\eta^2\end{array}\right)\,,\quad I+\La\Om=
\left(\begin{array}{cc}1+\La^1_1\eta^1 \bar\eta^1 &
\La^1_2\eta^2\bar\eta^2\\\La^2_1 \eta^1
\bar\eta^1&
1+\La^2_2\eta^2\bar\eta^2\end{array}\right)\,.$$
We easily compute
\begin{eqnarray*}\det(I+\La\Om) &=& 1+ \La^1_1\eta^1 \bar\eta^1 + 
\La^2_2\eta^2\bar\eta^2 +(\det\La)\eta^1 \bar\eta^1
\eta^2\bar\eta^2\,,\\
\det(I+\La\Om)\inv &=& 1 - \La^1_1\eta^1 \bar\eta^1 - 
\La^2_2\eta^2\bar\eta^2 + (2\La^1_1 \La^2_2-\det\La)\eta^1 \bar\eta^1
\eta^2\bar\eta^2\,,\\
\int \det(I+\La\Om)\inv d\bar\eta^1 d\eta^1 
d\bar\eta^2 d\eta^2 &=& 2\La^1_1 \La^2_2-\det\La \ =\ \La^1_1 \La^2_2+ \La^1_2
\La^2_1\,.\end{eqnarray*}

Hence by Theorem \ref{susy}, we have
\begin{equation} \kappa_{11}(r) = \frac{\La^1_1 \La^2_2+ \La^1_2
\La^2_1}{\det A}\,.\label{kappa11}\end{equation}
To obtain the explicit formula for $\kappa_{11}(r)$ \cite{Ha,BSZ1},  we set
$z^1=(r,0,\dots,0),\ z^2=0$ and substitute in (\ref{kappa11}) the resulting values
of $\La^p_{p'}$ and $\det A$  (see (\ref{sparse})--(\ref{detA}) below).

To obtain formulas for $\kappa_{km}(r)$ for higher $k,m$, we again set
$z^1=(r,0,\dots,0),\ z^2=0$. Using (\ref{heisenberg}), (\ref{ABC}) and
(\ref{La}), we see that
$\La^{pjq}_{p'j'q'}=0$ for $(j,q)\ne(j',q')$ and 
 \begin{equation}\label{sparse}
\left(\begin{array}{cc} \La^{1j1}_{1j1} & \La^{1j1}_{2j1}\\[8pt]
\La^{2j1}_{1j1}&
\La^{2j1}_{2j1}
\end{array}\right)=\left(\begin{array}{cc} P&Q\\Q&P
\end{array}\right)\,,\quad
\left(\begin{array}{cc} \La^{1jq}_{1jq} & \La^{1jq}_{2jq}\\[8pt]
\La^{2jq}_{1jq}&
\La^{2jq}_{2jq}
\end{array}\right)=\left(\begin{array}{cc}
R&S\\S&R \end{array}\right)\ \mbox{for}\ q\ge 2\,,\end{equation}
where
\begin{equation}\label{PQRS} \begin{array}{llllll}P&=&\displaystyle
{\frac {1-{e^{-{r}^{2}}}-{r}^{2}{e^{-{r}^{2}}}}{1-{e^{-{r}^{2}}}}}
\,, \quad &Q&=&\displaystyle
{\frac {{e^{-\half {r}^{2}}}\left (1-{e^{-{r}^{2}}}-{r}^{2}\right )}{1-
{e^{-{r}^{2}}}}}\,,\\[10pt]
R &=& 1\,, &S&=& e^{-\half {r}^{2}}\,.
\end{array}\end{equation}
We also have
$$A(r)=\left(\begin{array}{cc}1& e^{-\half r^2}\\e^{-\half r^2} &1
\end{array}\right)$$ and thus
\begin{equation}\label{detA} \det A = 1-e^{-r^2}\,.\end{equation}

We can write the $2km\times 2km$ matrices $\La,\Om$ in block form:
$$\La=\left(\begin{array}{cccc}\La' & 0
 &\cdots&0\\ 
0 &\La'' &
\cdots&0\\ 
\vdots&\vdots&\ddots&\vdots\\ 
0&0&\cdots&\La''\end{array}\right)\ ,\quad
\Om=\left(\begin{array}{cccc}\Om' & 0
 &\cdots&0\\ 
0 &\Om' &
\cdots&0\\
\vdots&\vdots&\ddots&\vdots\\ 
0&0&\cdots&\Om'\end{array}\right)\,,$$
$$\La'=\left(\begin{array}{cc}PI_k & QI_k\\ 
QI_k & PI_k\end{array}\right)\ ,\quad
\La''=\left(\begin{array}{cc}RI_k & SI_k\\ 
SI_k & RI_k\end{array}\right)\ ,\quad
\Om'=\left(\begin{array}{cc}\Om_1 & 0\\ 
0 & \Om_2\end{array}\right)\,,$$
where $I_k$ is the $k\times k$ identity matrix and
$$\Om_p=\left(\begin{array}{ccc}\eta^p_1\bar\eta^p_1 
 &\cdots&\eta^p_k\bar\eta^p_1\\ 
\vdots&&\vdots\\ 
\eta^p_1\bar\eta^p_k
 &\cdots&\eta^p_k\bar\eta^p_k
\end{array}\right)\,,\qquad p=1,2\,.$$

Hence 
\begin{equation}\label{Psi}{\det(I+\La\Om)} =
\Phi\Psi^{m-1}\,,\end{equation} where 
\begin{equation}\Phi = 
{\det(I+\La'\,\Om')}\,,\quad \Psi=
{\det(I+\La''\,\Om')}\,.\end{equation}
We note that
$$I+\La'\Om'=  \left(\begin{array}{cc}I+P\Om_1 & Q\Om_2\\
Q\Om_1 & I+P\Om_2\end{array}\right)\,,$$ and thus
\begin{eqnarray} \Phi &=&
\det(I+P\Om_1)\det\left[I+P\Om_2-Q^2\Om_1(I+P\Om_1)\inv\Om_2\right]\nonumber\\
&=& \det\left[I+P(\Om_1+\Om_2)+(P^2-Q^2)\Om_1\Om_2\right]\,.\end{eqnarray}
Similarly,
\begin{equation}\label{Psi2} \Psi = 
\det\left[I+R(\Om_1+\Om_2)+(R^2-S^2)\Om_1\Om_2\right]\,.\end{equation}
We recall that by Theorem \ref{susy}, \begin{equation}\label{s}
\kappa_{km}(r)=\frac{[(m-k)!]^2}{(m!)^2(1-e^{-r^2})^k}
\int\frac{1}{\det(I+\La\Om)}d\eta\,.\end{equation} Combining
(\ref{PQRS})--(\ref{s}), we obtain Corollary \ref{susy2}. \qed

\medskip To obtain explicit formulas for the pair correlation in a fixed
codimension
$k$ (for all dimensions $m$), we write $\Psi =1-(1-\Psi)$ and our
formula becomes:
\begin{equation}\kappa_{km}(r)=
\frac{[(m-k)!]^2}{(m!)^2(1-e^{-r^2})^k}\sum_{t=0}^{2k}
{m+t-2 \choose t}\int\Phi\inv (1-\Psi)^t\,d\eta\,.\label{km}\end{equation}
Using  Maple$^{\rm TM}$, we evaluate (\ref{km}) to obtain the following
pair correlation formulas:
\begin{eqnarray*}\kappa_{1m}(r)&=&\frac{1}{m^2\det A}\Big[{P}^{2}+2\left (m-1\right
)PR+{Q}^{2}+\left (m-1\right )^{2}{R}^{2} +\left (m-1\right ){S}^{2}\Big]\,,\\[5pt]
\kappa_{2m}(r)&=&\frac{1}{m^2(m-1)\det A^2}\Big[4(m-1){P}^{2}{R}^{2}+2{P}^{2
}{S}^{2}+4\left (m-1\right )\left (m-2\right
)P{R}^{3}\\&&\qquad +\ 4(m-2)PR{S}^{2}+2\left (m-1\right
){Q}^{2}{R}^{2}+4{Q}^{2}{S}^{ 2}+\left (m-1\right )\left (m-2\right
)^{2}{R}^{4}\\&&\qquad +\ 2\left (m-2
\right )^{2}{R}^{2}{S}^{2}+2\left (m-2\right ){S}^{4}\big]\,,\\[5pt]
\kappa_{3m}(r)&=&\frac{1}{m^2(m-1)(m-2)\det A^3}\Big [9\left (m-1\right
)\left  (m-2\right ){P}^{2}{R}^{4}+12\left (m-2\right ){P}^{2}{R}^{2}{S}^{
2}\\&&\qquad+\ 6{P}^{2}{S}^{4}+6\left (m-3\right )\left (m-1\right )\left (m-2
\right )P{R}^{5}+12\left (m-3\right )\left (m-2\right )P{R}^{3}{S}^{
2}\\&&\qquad+\ 12\left (m-3\right )PR{S}^{4}+3\left (m-1\right )\left (m-2
\right ){Q}^{2}{R}^{4}\\&&\qquad+\ 12\left (m-2\right ){Q}^{2}{R}^{2}{S}^{2}+
18{Q}^{2}{S}^{4}+\left (m-1\right )\left (m-2\right )\left (m-3
\right )^{2}{R}^{6}\\&&\qquad+\ 3\left (m-2\right )\left (m-3\right )^{2}{R}^{4}
{S}^{2}+6\left (m-3\right )^{2}{R}^{2}{S}^{4}+6\left (m-3\right 
){S}^{6}\Big ]\,.\end{eqnarray*}

Recalling (\ref{PQRS})--(\ref{detA}), we then obtain power
series expansions of the pair correlation function in codimensions $1,2,3$:

\begin{eqnarray*}\kappa_{1m}(r) &=&{\frac {m-1}{m}}{r}^{-2}+ {\frac {m-1}{2\,m}}+ {\frac
{\left (m+2\right )\left (m+1\right )}{12\,{m}^{2}}}{r}^{2} -{\frac {\left (m+4\right
)\left (m+3\right )}{720\,{m}^{2}}}{r}^{6}\\ && +{\frac {
\left (m+6\right )\left (m+5\right )}{30240{m}^{2}}}{r}^{10}
-{\frac {\left (m+8\right )
\left (m+7\right )}{1209600\,{m}^{2}}}{r}^{14} \cdots\;,\\[6pt]
\kappa_{2m}(r)&=&\frac{m-2}{m}\,r^{-4}+\frac{m-2}{m}\,r^{-2}
+\frac{5m^2-7m+12}{12(m-1)m}
+\frac{(m-2)(m+2)(m+1)}{12(m-1)m^2}\,r^2\\
&& +\frac{(m+3)(m+2)}{240(m-1)m}\,r^4
-\frac{(m-2)(m+4)(m+3)}{720(m-1)m^2}\,r^6+\dots\,,\\[6pt]
\kappa_{3m}(r) &=& {\frac {m-3}{m}}{r}^{-6} +
{\frac{{3}(m-3)}{2\,m}}{r}^{-4}
+{\frac {{m}^{2}-4\,m+6}{\left (m-2\right 
)m}}{r}^{-2}
+{
\frac {\left (m-3\right )\left (3\,{m}^{2}-m+8\right )}{8\,m\left (m-1
\right )\left (m-2\right )}}\\&&
+{\frac {
\left (m+2\right )\left (m+1\right )\left (19\,{m}^{2}-79\,m+120
\right )}{240\,{m}^{2}\left (m-1\right )\left (m-2\right )}}{r}^{2}
\\&&+{
\frac {\left (m-3\right )\left (m+3\right )\left (m+2\right )}{
160\,m\left (m-1\right )\left (m-2\right )}}{r}^{4}\cdots\;.\end{eqnarray*}

\medskip\noindent (The power series for $\kappa_{1m}$ and $\kappa_{2m}$ were given in
\cite{BSZ1} and \cite{BSZ2} respectively.) 

\section{The point case} \label{point}

We now prove Theorem \ref{2mm}: For the
case $k=m$, where the zero set
is discrete, 
(\ref{slpc}) becomes:

\begin{equation}\label{slpc-p1}  \kappa_{mm}(r) =
\frac{G_m(r)}{({m!})^2\det
A(r)^m}\,,
\end{equation} where
\begin{equation}\label{slpc-p2}G_m(r)=G_{2mm}(r)=
\int_{\C^{2m^2}}\left| 
\det_{1\le j,q\le m}\left(
 \xi_{jq}^1 
\right)
\det_{1\le j,q\le m}\left(
 \xi_{jq}^2 
\right)\right|^2d\ga_{\La(r)}(\xi)\,.\end{equation} 

We let $\big\langle \cdot\big\rangle_{\La(r)}=
\int\!\cdot d\ga_{\La(r)} $ denote the expected value with
respect to the Gaussian probability measure $\ga_{\La(r)}$. 
Thus  \begin{eqnarray} G_m(r) &=& \Big\langle \det(\xi^1)
\det(\xi^2)\det(\bar\xi^1)\det(\bar\xi^2)
\Big\rangle_{\La(r)}\nonumber\\
\label{slpc-p3}
&=&\sum_{\al,\be,\mu,\nu}(-1)^{\al+\be+\mu+\nu}
\left\langle\left(\prod_q\xi^1_{\al_qq}
\right)\left(\prod_{q}\xi^2_{\be_{q}q}\right)\left(\prod_{q}
\bar\xi^1_{\mu_{q}q}\right)\left(\prod_{q}
\bar\xi^2_{\nu_{q}q}\right)
\right\rangle_{\La(r)}\,,\end{eqnarray}
where the sum is over all 4-tuples $\al,\be,\mu,\nu\in\scal_m$ ($=$
permutations of
$\{1,\dots,m\}$), and $(-1)^\sigma$ stands for the sign of the
permutation $\sigma$.  We shall compute the terms of (\ref{slpc-p3}) using
the Wick formula rather than directly from the Berezin integral formula.  The computations simplifies considerably  since the
matrix
$\La(r)$ is sparse.  In fact, we shall see that the sign
$(-1)^{\al+\be+\mu+\nu}$ is positive whenever the corresponding moment is
nonzero.

Let us now  evaluate the moments
of order $4m$ in (\ref{slpc-p3}):
\begin{equation}\mcal_{\al\be\mu\nu}:=
\left\langle\left(\prod_q\xi^1_{\al_qq}
\right)\left(\prod_{q}\xi^2_{\be_{q}q}\right)\left(\prod_{q}
\bar\xi^1_{\mu_{q}q}\right)\left(\prod_{q}
\bar\xi^2_{\nu_{q}q}\right)
\right\rangle_{\La(r)}\,.\label{term}\end{equation}
Recall that the Wick formula expresses $\mcal_{\al\be\mu\nu}$ as a sum of
products of second moments with respect to the Gaussian measure
$\ga_{\La(r)}$.  Since this Gaussian is complex, these second
moments come from pairings of $\xi$'s with $\bar\xi$'s. 
We write \begin{equation}\label{Lambda}
\La_{p'j'q'}^{pjq}=\de_{j'}^j \La_{p'q'}^{pq} =
\left\langle\xi^p_{jq}\bar\xi^{p'}_{j'q'}
\right\rangle_{\La(r)}\,.\end{equation}
 Hence the term 
$\mcal_{\al\be\mu\nu}$ equals the permanent of the submatrix of
$\big (\La_{p'j'q'}^{pjq}\big)$ formed from the rows corresponding to
the variables 
$\xi^1_{\al_11},\dots,\xi^1_{\al_mm}, \xi^2_{\be_11} 
,\dots,\xi^2_{\be_mm}$ and columns corresponding to 
$\bar\xi^1_{\mu(1)1},\dots,\bar\xi^1_{\mu(m)m}, \bar\xi^2_{\nu(1)1} 
,\dots,\bar\xi^2_{\nu(m)m}$:
\begin{equation}\left(\begin{array}{ccc|ccc}\La^{1\al_11}_{1\mu_11} &
\cdots&\La^{1\al_11}_{1\mu_mm} &
\La^{1\al_11}_{2\nu_11} &\cdots&\La^{1\al_11}_{2\nu_mm}\\
\vdots &  &\vdots & \vdots &  &\vdots\\
\La^{1\al_mm}_{1\mu_11} &
\cdots&\La^{1\al_mm}_{1\mu_mm} &
\La^{1\al_mm}_{2\nu_11} &\cdots&\La^{1\al_mm}_{2\nu_mm}\\ \hline
\La^{2\be_11}_{1\mu_11} &
\cdots&\La^{2\be_11}_{1\mu_mm} &
\La^{2\be_11}_{2\nu_11} &\cdots&\La^{2\be_11}_{2\nu_mm}\\
\vdots &  &\vdots & \vdots &  &\vdots\\
\La^{2\be_mm}_{1\mu_11} &
\cdots&\La^{2\be_mm}_{1\mu_mm} &
\La^{2\be_mm}_{2\nu_11} &\cdots&\La^{2\be_mm}_{2\nu_mm}
\end{array}\right)\;.\label{bigmatrix}\end{equation}
(Recall that \ permanent$(V_{ij})=\sum_\sigma\prod_i V_{i\sigma_i}$,
where the sum is over all permutations $\sigma$.) 

To compute $\kappa_{mm}(r)$, we can set $z^1=(r,0,\dots,0),\
z^2=0$, as before.  Recalling (\ref{sparse}) and the fact that
$\La^{pjq}_{p'j'q'}=0$ for $(j,q)\ne(j',q')$, we observe that
(\ref{bigmatrix}) is made up of 4 diagonal matrices. For example, if
$m=3$, then (\ref{bigmatrix}) becomes

$$\left(\begin{array}{ccc|ccc}\de^{\al_1}_{\mu_1}P & 0&0 &
\de^{\al_1}_{\nu_1}Q &0&0\\
0 & \de^{\al_2}_{\mu_2}R&0&0 &\de^{\al_2}_{\nu_2}S&0\\
0&0& \de^{\al_3}_{\mu_3}R &0&0& \de^{\al_3}_{\nu_3}S\\ \hline
\de^{\be_1}_{\mu_1}Q & 0&0 &
\de^{\be_1}_{\nu_1}P &0&0\\
0 & \de^{\be_2}_{\mu_2}S&0&0 &\de^{\be_2}_{\nu_2}R&0\\
0&0& \de^{\be_3}_{\mu_3}S &0&0& \de^{\be_3}_{\nu_3}R\end{array}\right)\;.$$

\medskip\noindent We conclude that $\mcal_{\al\be\mu\nu}$ is a product of
$2\times 2$ permanents:
\begin{equation}\label{product}\mcal_{\al\be\mu\nu}=\left(\de^{\al_1}_{\mu_1}
\de^{\be_1}_{\nu_1}P^2 + \de^{\al_1}_{\nu_1} \de^{\be_1}_{\mu_1}Q^2\right)
\prod_{q=2}^m \left(\de^{\al_q}_{\mu_q}
\de^{\be_q}_{\nu_q}R^2 + \de^{\al_q}_{\nu_q} \de^{\be_q}_{\mu_q}S^2\right)
\,.\end{equation}  
In particular, $\mcal_{\al\be\mu\nu}$ vanishes unless
\begin{equation}\{\mu_q,\nu
_q\}=\{\al_q,\be_q\}\qquad \mbox{for}\ 1\le\ q\le
m\,.\label{nv}\end{equation}  We now claim that (\ref{nv}) implies that 
$(-1)^{\al+\be+\mu+\nu}=1$: First of all, by multiplying the four
permutations by $\al\inv$ on the left, we can assume without loss of
generality that $\alpha_i=i$ for all $i$.  Now write $\beta$ as a product of
disjoint cycles.  Then one sees that $\mu$ is a product of some of
these cycles and $\nu$ is a product of the other cycles, and the
positivity of the product of signs easily follows.  

Hence equation (\ref{slpc-p3}) becomes
\begin{equation}\label{slpc-p4} G_m(r)=
\sum_{\al,\be,\mu,\nu}\mcal_{\al\be\mu\nu}\,.\end{equation}
We now use (\ref{slpc-p3}) to evaluate $G_m$ for arbitrary dimension $m$.

\begin{lem} $\quad G_m= (m-1)!m!\big[P^2 f_m(R^2,S^2) +Q^2
f_m(S^2,R^2)\big]$, 

\noindent where
\begin{eqnarray*}f_m(x,y)&=& y^{m-1} + 2 x y^{m-2} + \cdots + (m-1) x^{m-2} y
+ m x^{m-1}\\& =& {\frac {m{x}^{m+1}+{y}^{m+1}-\left (m+1\right
){x}^{m}y}{\left (x-y
\right )^{2}}}\,.\end{eqnarray*} \label{Gm}\end{lem}

\begin{proof} We use induction on $m$.  The identity holds trivially for
$m=1$, since $f_1=1$ and $G_1 =P^2+Q^2$. Let $m\ge2$ and suppose the
identity has been verified for $1,\dots,m-1$.  Since 
$\mcal_{\al\be\mu\nu}$ is unchanged if we multiply all the permutations on
the left by $\al\inv$, we have $G_m
=m!G^0_m$, where $G_m^0=\sum _{\be,\mu,\nu}\mcal_{e\be\mu\nu}$ ($e=\
$identity). 

For $1\le i\le m$, let $\ccal_i\subset\scal_m$ denote the collection of
$i$-cycles of the form 
$(1a_2\dots a_i)$.  For each $\sigma\in\ccal_i$, let $\sigma^\perp$ denote
those permutations $\tau\in \scal_m$ that fix the elements
$1,a_1,\dots,a_i$.  We
claim that 
\begin{equation}\label{RS}\sum_{\be\in\sigma^\perp} 
\sum_{\mu,\nu}\mcal_{e\be\mu\nu}=(m-i)!(P^2R^{2i-2}+Q^2S^{2i-2})
g_{m-i}(R^2,S^2)\,,\end{equation}
where $g_l(x,y)=x^l+x^{l-1}y+\cdots+xy^{l-1}+y^l$.

To verify (\ref{RS}), we can assume without loss of generality that
$\sigma=(1\dots i)$.  Recall that we need only consider permutations $\mu$
that are products of some of the cycles in $\be$ (and $\nu$ is determined
by the pair $(\be,\mu)$, since $\nu$ is the product of the other cycles of
$\be$ when $\mcal_{e\be\mu\nu}\ne 0$).  For the $P^2$-terms of the sum, $\nu$
contains the cycle $(1\dots i)$ so that $\mu_q=q,\ \nu_q=\sigma_q=\be_q$
for
$q=1,\dots,i$. (For the
$Q^2$-terms,
$\mu$ contains $(1\dots i)$.)  Hence by (\ref{product}), we have
\begin{equation}\label{prod2}
\sum_{\be\in\sigma^\perp}\sum_{\mu,\nu}\mcal_{e\be\mu\nu}=P^2 R^{2i-2} 
\sum_{\be\in\sigma^\perp}{\sum_{\mu,\nu}}'\prod_{q=i+1}^m
\left(\de^{q}_{\mu_q}
\de^{\be_q}_{\nu_q}R^2 + \de^{q}_{\nu_q} \de^{\be_q}_{\mu_q}S^2\right)
+ \mbox{terms with}\ Q^2\,,\end{equation} where $\sum'$ is over those $\mu
,\nu$ with $\mu_q=q,\ \nu_q=\be_q$, for $1\le q\le i$. To compute the double
sum in the right side of  (\ref{prod2}), we notice by (\ref{product}) and
(\ref{slpc-p4}) that it equals $\frac{1}{(m-i)!}G_{m-i}$ with $P,Q$ replaced
by $R,S$ respectively.  Hence by our inductive assumption, we have 
\begin{eqnarray*}\sum_{\be\in\sigma^\perp}{\sum_{\mu,\nu}}'\prod_{q=i+1}^m
\left(\de^{q}_{\mu_q}
\de^{\be_q}_{\nu_q}R^2 + \de^{q}_{\nu_q} \de^{\be_q}_{\mu_q}S^2\right)
&=& (m-i-1)![R^2f_{m-i}(R^2,S^2)+S^2 f_{m-i}(S^2,R^2)]\\
&=& (m-i)!g_{m-i}(R^2,S^2)\,.\end{eqnarray*}
The computation of the $Q^2$ terms is similar, and hence (\ref{RS}) holds.

We now have by (\ref{RS}),
$$G_m=m!\sum_{i=1}^m\sum_{\sigma\in\ccal_i}\sum_{\be\in\sigma^\perp} 
\sum_{\mu,\nu}\mcal_{e\be\mu\nu} = m!\sum_{i=1}^m (\#\ccal_i)
(m-i)!(P^2R^{2i-2}+Q^2S^{2i-2})
g_{m-i}(R^2,S^2)\,.$$ Noting that $(\#\ccal_i)(m-i)!=(m-1)!$ and summing
over $i$, we obtain the desired formula. \end{proof}

We now complete the proof of Theorem \ref{2mm}.
By (\ref{slpc-p1}) and Lemma \ref{Gm}, we have
\begin{equation}\kappa_{mm}(r)= \frac{P^2 f_m(R^2,S^2) +Q^2
f_m(S^2,R^2)}{m(\det A(r))^m}\,.\label{kmm-}\end{equation}
Substituting (\ref{PQRS})--(\ref{detA}) into
(\ref{kmm-}), we obtain (\ref{pointpaircor}). 

Finally, to verify (\ref{leading}), we note by (\ref{PQRS}) that
$P=\half r^2+\cdots,\
Q=\half r^2+\cdots,\ R=1,\ S=1+\cdots$, and hence 
$$\kappa_{mm}=\frac {2f_m(1,1)(r^4/4)+\cdots}{m
r^{2m}+\cdots}=\frac{f_m(1,1)}{2m} r^{4-2m} +\cdots = \frac{m+1}{4}
 r^{4-2m} +\cdots\,.$$
The following proposition yields the remainder estimate of (\ref{leading}).

\begin{prop} If $m$ is odd, resp.\ even, then $\kappa_{mm}(r)$ is an
odd,  resp.\ even, function of $r^2$. \label{oddeven}\end{prop}

\begin{proof}  Let $\wh P,\wh Q$ be the functions given by $P(r)=\wh P(u),\ 
Q(r)=\wh Q(u)$, where $u=r^2$.  From (\ref{kmm-}), we have

$$\kappa_{mm}= \frac{\wh P(u)^2 f_m(1,e^{-u}) +\wh Q(u)^2
f_m(e^{-u},1)}{m(1-e^{-u})^m}\,.$$
We observe from (\ref{PQRS}) that $\wh P(-u) = e^{u/2}\wh Q(u)$ and thus
$$\kappa_{mm}(-u) = \frac{e^u \wh Q(u)^2 f_m(1,e^{u}) + e^u
\wh P(u)^2 f_m(e^{u},1)}{m(1-e^{u})^m} = (-1)^m\kappa_{mm}(u)\,,$$
since $f_m$ is homogeneous of order $m-1$.\end{proof}

The expansions of (\ref{pointpaircor}) are easily obtained using
Maple$^{\rm TM}$:
\begin{eqnarray*}\kappa_{11}(r)&=&\half\,{r}^{2}-\frac{1}{36}\,{r}^{6}+{\frac
{1}{720}}\,{r}^{10}-{\frac {1}{16800 }}\,{r}^{14}+{\frac
{1}{435456}}\,{r}^{18}-{\frac {691}{8382528000}}\, {r}^{22}\ \cdots\\
\kappa_{22}(r)&=& \frac{3}{4}+\frac{1}{24}\,{r}^{4}- {\frac
{1}{288}}\,{r}^{8}+{\frac {1}{4800}}\,{r}^{12 }-{\frac
{1}{96768}}\,{r}^{16}+{\frac {691}{1524096000}}\,{r}^{20} \ \cdots\\
\kappa_{33}(r)&=&
{r}^{-2}+\frac{1}{4}\,{r}^{2}-{\frac {11}{2160}}\,{r}^{6}-{\frac
{1}{50400}}\, {r}^{10}+{\frac {1}{80640}}\,{r}^{14}-{\frac
{4871}{5029516800}}\,{r}^ {18}\ \cdots\\
\kappa_{44}(r)&=& \frac
{5}{4}\,{r}^{-4}+{\frac {95}{144}}+{\frac {19}{576}}\,{r}^{4}-{\frac {79}
{40320}}\,{r}^{8}+{\frac {7}{82944}}\,{r}^{12}-{\frac {6049}{
2235340800}}\,{r}^{16}\ \cdots\\
\kappa_{55}(r)&=& \frac
{3}{2}\,{r}^{-6}+\frac{4}{3}\,{r}^{-2}+{\frac {55}{288}}\,{r}^{2}-{\frac
{19}{ 16800}}\,{r}^{6}-{\frac {257}{1451520}}\,{r}^{10}+{\frac {21337}{
1397088000}}\,{r}^{14} \ \cdots\\
\kappa_{66}(r)&=& \frac
{7}{4}\,{r}^{-8}+\frac{7}{3}\,{r}^{-4}+{\frac {5257}{8640}}+{\frac
{407}{14400}}
\,{r}^{4}-{\frac {103}{82944}}\,{r}^{8}+{\frac {38177}{1197504000}}\,{
r}^{12}\ \cdots
\end{eqnarray*}

\end{document}